\documentclass[prb,a4paper,twocolumn,showpacs,superscriptaddress]{revtex4}

\usepackage{amsmath}
\usepackage{amssymb}
\usepackage{graphicx}
\usepackage{bm}

\newcommand{\bfsfG}{\mbox{\sffamily\bfseries{G}}}
\newcommand{\sfG}{\mbox{\sffamily{G}}}

\begin{document}

\title{Hyperbolic metamaterials: nonlocal response regularizes broadband super-singularity}
\author{Wei Yan, Martijn Wubs, and N. Asger  Mortensen\footnote{asger@mailaps.org}}
\address{DTU Fotonik, Department of Photonics Engineering, Technical University of Denmark, DK-2800 Kongens Lyngby, Denmark}

\date{\today}

\begin{abstract}
We study metamaterials known as hyperbolic media that in the usual local-response approximation exhibit hyperbolic dispersion and an associated broadband singularity in the density of states. Instead, from the more microscopic hydrodynamic Drude theory we derive qualitatively different optical properties of these metamaterials, due to the free-electron nonlocal optical response of their metal constituents. We demonstrate that nonlocal response gives rise to a large-wavevector cutoff in the dispersion that is inversely proportional to the Fermi velocity of the electron gas, but also for small wavevectors we find differences for the hyperbolic dispersion. Moreover, the size of the unit cell influences effective parameters of the metamaterial even in the deep sub-wavelength regime. Finally, instead of the broadband super-singularity in the local density of states, we predict a large but finite maximal enhancement proportional to the inverse cube of the Fermi velocity.
\end{abstract}

\maketitle

\section{Introduction}

Metamaterials, consisting of subwavelength artificial unit cells, show a great potential in optical applications, such as perfect lenses~\cite{Pendry:2000} and invisibility cloaks.\cite{Pendry:2006,Ulf:2006} Hyperbolic metamaterials (HMM) 
are of special interest because of their unusual hyperbolic dispersion curves~\cite{Smith:03,Smolyaninov:2011,Jacob:2009,Jacob:2010,Poddubny:2011,Noginov:2010,Tumkur:2011,Kidwai:2011,Noginov:2009,Yao:2008,Yan:2009} that support radiative modes with unbounded wavenumbers. Because of the diverging radiative local density of states (LDOS), a point emitter in such a medium would exhibit instantaneous radiative decay.\cite{Jacob:2009,Jacob:2010,Poddubny:2011,Noginov:2010,Tumkur:2011,Kidwai:2011,Noginov:2009} This broadband `super-singularity'~\cite{Jacob:2009,Noginov:2010} is indeed broadband, since the hyperbolic dispersion does not rely on specific resonances. The prediction by Jacob {\em et al.}~\cite{Jacob:2009} that hyperbolic media thereby form a new route to enhanced light-matter coupling recently found experimental support.\cite{Noginov:2010,Tumkur:2011}
These measured lifetimes were nonzero, and state-of-the-art theories explain this from three parameters: the nonvanishing damping~$\gamma$,  the size~$a$ of the unit cell, and  the size~$D$ of the emitter.\cite{Jacob:2009,Jacob:2010,Poddubny:2011} In particular, as a function of these parameters the radiative LDOS scales as $\gamma^{-3/2}$~(Ref.~\onlinecite{Jacob:2009}), $a^{-3}$~(Ref.~\onlinecite{Jacob:2010}) and $D^{-3}$~(Ref.~\onlinecite{Poddubny:2011}), respectively.
Usually unit cells are larger than emitter sizes, which makes $a$ the more important limiting factor to the radiative LDOS.


%
\begin{figure}[h!]
\includegraphics[width=0.5\textwidth]{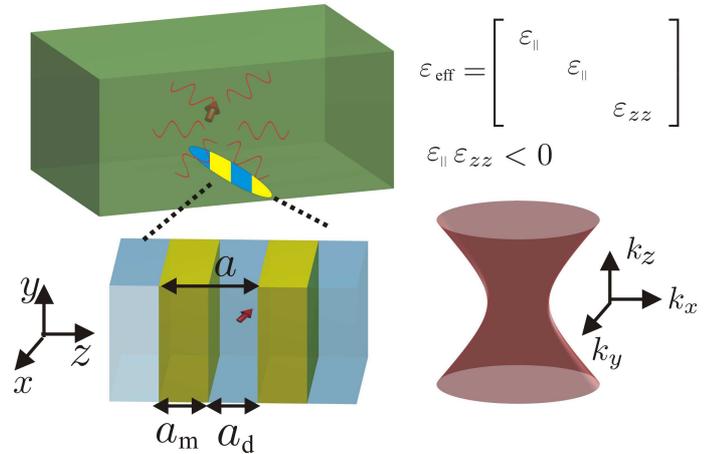}
\caption{(Color online) Sketch of a multilayer hyperbolic metamaterial consisting of periodic dielectric-metal bilayers.  The dielectric and metal layer thicknesses are $a_{\rm d}$ and $a_{\rm m}$, respectively, and their sum equals the period $a$ of the unit cell. Corresponding permittivities are $\epsilon_{\rm d}$ and $\epsilon_{\rm m}$. The red arrow is a dipole emitter located in the middle of a dielectric layer.}
\label{fig1}
\end{figure}

Owing to the great recent progress in nano-fabrication techniques, the scale on which metamaterials can be patterned is entering the nanometer regime, where nonlocal response of the metal becomes important.\cite{Bloch:1933,Boardman:1982,Raza:2011,David:2011,Abajo:2008,Toscano:1,Toscano:2,FernandezDominguez:2012,Ciraci:2012a,Marier:2012,Ciraci:2012b} For example, nonlocal response can significantly blueshift the localized surface plasmon polariton (SPP) resonance peak and modify the field enhancement of a nanoscale plasmonic structure.\cite{David:2011,Abajo:2008,Toscano:1,Toscano:2,FernandezDominguez:2012,Scholl:2012}

In this paper, we discuss the effects of nonlocal response on the optical properties of hyperbolic metamaterials. It is shown that the nonlocal response gives rise a large-wavevector cutoff in the dispersion, inversely proportional to the Fermi velocity of the electron gas. In fact, the dispersion in hyperbolic media becomes no longer strictly hyperbolic. Accordingly, we identify a new and fundamental limit on the enhancement of the radiative emission rates of HMMs. In particular, we show that the radiative LDOS does not grow arbitrarily large even in the ideal limiting case that all three aforementioned parameters~$\gamma$, $a$, and $D$ vanish, since the intrinsic nonlocal response turns the `super-singularity' into a finite broadband LDOS enhancement.  On a more general level, our results illustrate the need, as for metallic nanoparticles~\cite{Scholl:2012,Ciraci:2012b}, to take nonlocal response into account in homogenization theories, where the goal is to predict the effective properties of metamaterials with ever decreasing unit cell sizes.

The paper is organized as follows: In Sec.~II, we introduce the linearized hydrodynamic Drude model within the Thomas--Fermi approximation. In Sec.~III, the unusual dispersion curves of the HMMs and their effective parameters are discussed. In Sec.~IV, to understand better the HMM dispersion found in Sec.~III, we discuss the SPP supported by a single metal layer. In Sec.~V, we investigate the LDOS of the HMMs, before discussing our results and concluding in Sec.~VI. Finally, details of the calculations can be found in Appendices~{A-C}.

\section{Hydrodynamic Drude model}
We consider a similar multilayer HMM geometry as in the recent experiments by Tumkur {\em et al.}~\cite{Tumkur:2011}, see Fig.~\ref{fig1}.
The unit cell is a sub-wavelength dielectric-metal bilayer, which is relatively simple and cheap to fabricate~\cite{Tumkur:2011} and allows analytical analysis. For an effective-medium description of such a metamaterial, a local-response approximation (LRA) is usually employed, i.e.  spatial dispersion
is neglected. This gives the  effective dispersion relation
\begin{equation}
\frac{k_z^2}{\epsilon_{zz}^{\rm{loc}}}+\frac{k_{\scriptscriptstyle \parallel}^2}{\epsilon_{\scriptscriptstyle\parallel}^{\rm{loc}}} =\frac{\omega^2}{c^2},
\end{equation}
where $\epsilon_{zz}^{\rm{loc}}=a (a_{\rm d}/\epsilon_{\rm d} +a_{\rm m}/\epsilon_{\rm m})^{- 1}$, $a \epsilon_{\scriptscriptstyle\parallel}^{\rm{loc}}= a_{\rm d} \epsilon_{\rm d} + a_{\rm m} \epsilon_{\rm m}$, and $k_{\scriptscriptstyle\parallel}=(k_x^2+k_y^2)^{1/2}$. Below the plasma frequency $\omega_{\rm p}$, where $\epsilon_{\rm m}<0$ in the Drude model of a pure plasma, the dielectric tensor elements $\epsilon_{zz}$ and $\epsilon_{\scriptscriptstyle\parallel}$ can have opposite signs by a proper choice of the filling factor $a_{\rm m}/a$. Then the dispersion becomes hyperbolic, meaning that an iso-frequency contour becomes a hyperbola rather than the usual ellipse in the $(k_{z},k_{\scriptscriptstyle\parallel})$-plane. The length of this contour diverges, and so does the radiative LDOS. Although we discuss metal-dielectric bilayer structures, we want to point out that our theory may also be applied to structures where the metal is replaced by other materials with a Drude response.\cite{Alekseyev:2012a}

In the present paper, we go beyond the LRA, and discuss the optical properties of the HMMs in the linearized hydrodynamic Drude model (HDM) within the Thomas-Fermi approximation.\cite{Bloch:1933,Boardman:1982,Raza:2011} In the HDM, the metal supports both the usual divergence-free (`transverse') and rotation-free (`longitudinal') waves. Above the plasma frequency both types of waves can propagate. The dispersion $k_{\rm T}(\omega)$ of the transverse waves is given by $\epsilon_{\rm m}^{\rm T}(\omega)\omega^{2} = k^{2}c^{2}$ while $k_{\rm L}(\omega)$ of the longitudinal waves follows from $\varepsilon_{m}^{\rm L}(k,\omega)=0$, in terms of the dielectric functions
\begin{subequations}\label{hdm}
\begin{eqnarray}
\epsilon_{\rm m}^{\rm T}(\omega) &=& 1 - \frac{{\omega_{\rm p}^2}}{{{\omega ^2} + i\omega\gamma}}\\
\epsilon_{\rm m}^{\rm L}(k,\omega)&=&1-\frac{\omega_{\rm p}^2}{\omega^2+i\omega\gamma-\beta^2k^2}.
\end{eqnarray}
\end{subequations}
Here, $\gamma$ is the Drude damping, $\omega_{\rm p}$ is the plasma frequency, and the nonlocal parameter $\beta$ is equal to $\sqrt{3/5}v_{\rm F}$ with $v_{\rm F}$ representing the Fermi-velocity. While $\epsilon_{\rm m}^{\rm T}$ is the familiar Drude dielectric function, $\epsilon_{\rm m}^{\rm L}$ depends on $v_{\rm F}$ and describes nonlocal response.

%

\section{Dispersion and effective material parameters}
To calculate the exact dispersion equation for the infinitely extended HMM, we employ a transfer-matrix method for both transverse and longitudinal waves combined. Our method is quite similar to the one developed by Moch\'an {\em et al.}~\cite{Mochan:1987}, but we corrected the additional boundary condition (ABC) that in Ref.~\onlinecite{Mochan:1987} was employed for simplicity. An ABC is required to complement the usual Maxwell boundary conditions, and all boundary conditions together make the solution to the coupled Maxwell and hydrodynamic equations unique. Details how to derive the correct ABC and a consistency check can be found in Appendix~A.


For arbitrary unit cell size $a$ and metal and dielectric filling fractions, we find the exact dispersion relation for the infinite HMM to be
\begin{widetext}
\begin{eqnarray}
\cos\theta_b&=&\Big\{ \cos {\theta _d}\Big[k_{{\rm L}z}\cos {\theta _m}\sin {\theta _l} - k_{\scriptscriptstyle \parallel}\frac{{({w_d} - {w_m})}}{{{z_m}}}\sin {\theta _m}\cos {\theta _l}\Big] +\sin {\theta _d}\Big[k_{\scriptscriptstyle \parallel}\frac{{({w_d} - {w_m})}}{{{z_d}}}(1 - \cos {\theta _m}\cos {\theta _l})\nonumber\\
&-&\frac{1}{2}\Big[\frac{k_{\scriptscriptstyle \parallel}^2}{k_{{\rm L}z}}\frac{{{}{{({w_d} - {w_m})}^2}}}{{{z_d}{z_m}}} + k_{{\rm L}z}\left(\tfrac{{{z_d}}}{{{z_m}}} + \tfrac{{{z_m}}}{{{z_d}}}\right)\Big]\sin {\theta _m}\sin {\theta _l}\Big]\Big\} {\Big[k_{{\rm L}z}\sin {\theta _l} - k_{\scriptscriptstyle \parallel}\frac{{({w_d} - {w_m})}}{{{z_m}}}\sin {\theta _m}\Big]^{ - 1}},\label{exact_dispersion}
\end{eqnarray}
\end{widetext}
with
\begin{eqnarray}
\theta_b&=&k_{\rm z}a,\;\theta_d=k_{\rm dz}a_{\rm d},\;\theta_m=k_{\rm mz}^{\rm T}a_{\rm m}\;\theta_l=k_{\rm mz}^{\rm L}a_{\rm m},\\
z_{\rm d}&=&\frac{k_{\rm dz}}{k_0\epsilon_{\rm d}},\;w_{\rm d}=\frac{k_{\scriptscriptstyle\parallel}}{k_0},\;z_{\rm m}=\frac{k_{\rm mz}^{\rm T}}{k_0\epsilon_{\rm m}^{\rm T}},\;w_{\rm d}=\frac{k_{\scriptscriptstyle\parallel}}{k_0\epsilon_{\rm m}^{\rm T}},
\end{eqnarray}
where $k_{\rm dz}^2+k_{\scriptscriptstyle\parallel}^2=\omega^2\epsilon_{\rm d}/c^2$, $(k_{\rm mz}^{\rm T})^2+k_{\scriptscriptstyle\parallel}^2=\omega^2\epsilon_{\rm m}^{\rm T}/c^2$, and $(k_{\rm mz}^{\rm L})^2+k_{\scriptscriptstyle\parallel}^2=k_{\rm L}^2$ with $k_{\rm L}^2=(\omega^2+i\gamma\omega-\omega_p^2)\beta^2$.
This dispersion equation looks very similar to the one found in Ref.~\onlinecite{Mochan:1987}, but the essential difference is that the parameter $w_{\rm d}$ here is $w_{\rm d}/\epsilon_d$ in Ref.~\onlinecite{Mochan:1987}.

\begin{figure}[h!]
\includegraphics[width=0.45\textwidth]{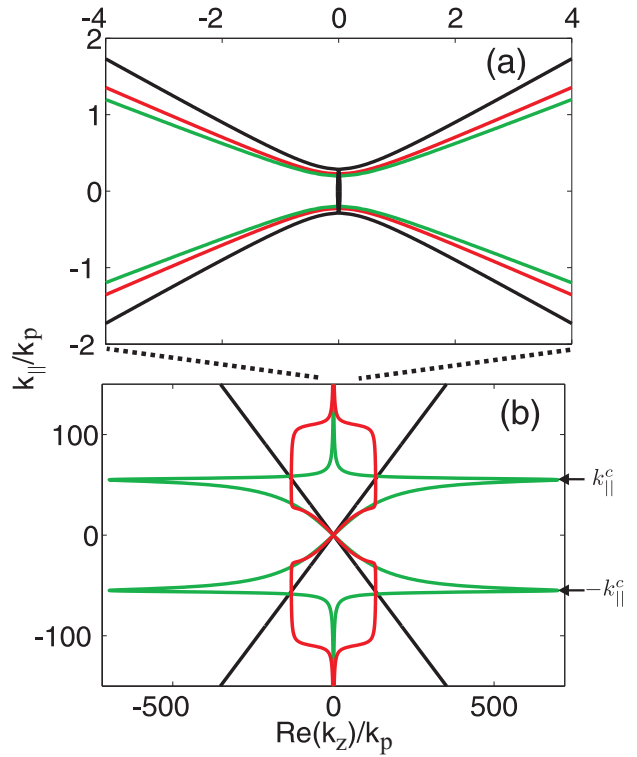}
\caption
{(Color online) Dispersion curves of the HMM for $\omega=0.2\omega_{\rm p}$, on  (a) small and (b) large wavevector intervals. Red curves for $a=\lambda_{\rm F}$, green curves for $a\to0$, black curves for $a\to0$ in the LRA. The unit cell of the HMM is a free-space-Au bilayer with $a_{\rm d}=a_{\rm m}=a/2$. Material parameters for Au: $\hbar {\omega_{\rm p}} = 8.812{\kern 1pt} \rm{eV}$, $\hbar {\gamma} = 0.0752{\kern 1pt} \rm{eV}$, and $v_{\rm F}=1.39\times10^6\rm{m/s}$.}
\label{fig2}
\end{figure}

As an example we consider a HMM with free-space-Au bilayer unit cell with $a_{\rm d}=a_{\rm m}=a/2$, and we include the Au Drude loss. Fig.~\ref{fig2} depicts HMM dispersion curves at $\omega=0.2\omega_{\rm p}$. Fig.~\ref{fig2}(a) shows hyperbolic dispersion in the small wavevector regime, whereas Fig.~\ref{fig2}(b) zooms out and shows strong deviations from hyperbolic dispersion for large wavevectors.

First, in Fig.~\ref{fig2}(a) we observe three non-coinciding hyperbolic dispersion curves in the small-$k$ region, one for local and two for nonlocal response. This tells us that the HDM is not just a local theory with a large-wave vector cutoff added, since then the curves for local and for nonlocal response would have coincided for small wavevectors. Furthermore, the two nonlocal hyperbolic curves do not coincide, the one for a strongly subwavelength unit cell $a = \lambda_{\rm F}$ and the other for $a \to 0$. This illustrates that the size of the unit cell affects effective-medium properties, even in the deep subwavelength limit, which goes against common wisdom obtained in the LRA. The reason for this is that in the HDM the longitudinal wave in the metal layer has a large vector $k_{\rm L}(\omega) \gg 2\pi/\lambda$, so that typically the condition $|k_{\rm L}(\omega)|a \ll 1$ is not satisfied even in the deep subwavelength limit. Thus, the longitudinal wave can probe the finite size of the unit cell even though $a \ll \lambda$, and this gives rise to the periodicity-dependent dispersion curve of Fig.~\ref{fig2}(a).

Zooming out, Fig.~\ref{fig2}(b) shows that nonlocal response gives rise to closed non-hyperbolic dispersion curves, for both considered values of $a$, in stark contrast to the familiar hyperbolic curve in the LRA which is also shown. (We still call these media hyperbolic because of their hyperbolic small-wavevector dispersion.) Both $k_{\scriptscriptstyle \parallel}$ and $k_{z}$ are bounded on the curve for $a=\lambda_{\rm F}$.
For smaller values of $a$, we do not expect the hydrodynamic Drude model to apply~\cite{Raza:2011,FernandezDominguez:2012}, but as we shall see below it is useful to also consider the limit $a\to 0$. The curve for $a\to 0$ shows a turning point at $k_{\scriptscriptstyle \parallel}=k_{\scriptscriptstyle \parallel}^c$. In the lossless limit,
no radiative modes exist above $k_{\scriptscriptstyle \parallel}^c$, as we explain shortly. The wavevector $k_{\scriptscriptstyle \parallel}^c$ is found to be
\begin{equation}
k_{\scriptscriptstyle \parallel}^c = \frac{\omega}{\beta}\propto \frac{\omega }{{{v_{\rm F}}}}.
\end{equation}

To analyze the dispersion curves of Fig.~\ref{fig2}, we derive the effective material parameters of the HMM  by a mean-field theory that can be applied to many geometries. In the limit of vanishing unit-cell size, we obtain the effective material parameters
\begin{equation}
{\epsilon_{zz}^{\rm{nloc}}}=\epsilon_{zz}^{\rm{d}},\;\;\;\epsilon_{\scriptscriptstyle \parallel}^{\rm{nloc}} = \epsilon_{\scriptscriptstyle \parallel}^{\rm{d}}\frac{{k_L^2\epsilon_{\scriptscriptstyle \parallel}^{\rm{loc}}/\epsilon_{\scriptscriptstyle \parallel}^{\rm{d}} - k_{\scriptscriptstyle \parallel}^2{\epsilon_{\rm m}^{\rm T}}}}{{k_L^2 - k_{\scriptscriptstyle \parallel}^2{\epsilon_{\rm m}^{\rm T}}}},
\label{eq4}
\end{equation}
where $\epsilon_{zz}^{\rm{d}}$ and $\epsilon_{\scriptscriptstyle\parallel}^{\rm{d}}$ represent the effective parameters of the metamaterials when the metal layer is replaced by a free-space layer, with $\epsilon_{zz}^{\rm{d}}=a (a_{\rm d}/\epsilon_{\rm d} +a_{\rm m})^{- 1}$, and $a \epsilon_{\scriptscriptstyle\parallel}^{\rm{d}}= a_{\rm d} \epsilon_{\rm d} + a_{\rm m}$. Both nonlocal effective material parameters of Eq.~(\ref{eq4}) differ from the corresponding parameters for local response. The derivations leading to Eq.~(\ref{eq4}) are presented in Appendix~B. Neglecting loss at first, we find from Eq.~(\ref{eq4}) that $\epsilon_{\scriptscriptstyle \parallel}^{\rm{nloc}}$ has a resonance at $k_{\scriptscriptstyle \parallel}=k_{\scriptscriptstyle \parallel}^c$ where both $\epsilon_{\scriptscriptstyle \parallel}^{\rm{nloc}}$ and $k_z$ diverge. The value of $k_{\scriptscriptstyle \parallel}^c$ is independent of $\epsilon_{\rm d}$ (unlike what one would find when using the incorrect ABC of Refs.~\cite{Mochan:1987,Ciraci:2012b}). Increasing  $k_{\scriptscriptstyle \parallel}$ beyond $k_{\scriptscriptstyle \parallel}^c$,  the $\epsilon_{\scriptscriptstyle \parallel}^{\rm{nloc}}$ changes sign from negative to positive. Since  ${\epsilon_{zz}^{\rm{nloc}}}$ is always positive, it follows that no mode exists above $k_{\scriptscriptstyle \parallel}^c$. Thus, nonlocal response gives rise to a large-wavenumber cutoff at $k_{\scriptscriptstyle \parallel}=k_{\scriptscriptstyle \parallel}^c$. With loss, the resonance is smoothed out and modes exist also above $k_{\scriptscriptstyle \parallel}^c$. However, for $k_{\scriptscriptstyle \parallel}\to\infty$, the corresponding $k_z$ approaches $i\infty$, which shows that such large-wavevector modes are purely evanescent. This explains why the dispersion curves in Fig.~\ref{fig2} are closed.

We stated in Eq.~(\ref{eq4}) that unlike in the LRA, in the HDM the effective parameter ${\epsilon_{zz}^{\rm{nloc}}}$ simply equals the (positive) permittivity $\epsilon_{zz}^{\rm{d}}$. This outcome is fixed for $a \to 0$ by the continuity of the normal components of the displacement field and the ABC of Eq.~(\ref{ABC1}) with $\epsilon_{\rm}^{\rm other}=1$.\cite{Raza:2011} In particular, the different boundary conditions explain why the local and nonlocal  $a \to 0$ curves in Fig.~\ref{fig2}(a) exhibit different hyperbolic  small-wavevector dispersion.

Above the plasma frequency, the HDM and the LRA also exhibit qualitatively different dispersion. In the LRA no hyperbolic dispersion exists for frequencies above the  plasma frequency, not even for small wavevectors, since then both $\epsilon_{\rm d}$ and $\epsilon_{\rm m}$ are positive. By contrast, hyperbolic dispersion can exist in the HDM for $\omega>\omega_{\rm p}$, because the effective-medium parameter $\epsilon_{\scriptscriptstyle \parallel}^{\rm{nloc}}$ given in Eq.~(\ref{eq4}) can assume negative values above $\omega_{\rm p}$.

\section{Surface plasmon polariton supported by a single metal layer}
In Sec.~III, it was demonstrated the dispersion curves in the LRA and HDM differ significantly. To understand this better, here we relate these essential differences to the different properties of single metal layers in both theories, knowing that the bulk modes of the HMM result from the coupling of SPPs of neighboring metal layers. So we investigate the SPPs supported by a single metal layer, first analytically in the quasi-static limit.
With respect to the magnetic field, the SPPs can be classified as even and odd modes. In the HDM, the dispersion relations of the even and odd modes are found to be
\begin{subequations}
\begin{eqnarray}
\tanh\left(\tfrac{{{k_{\rm sp}}{a_{\rm m}}}}{2}\right) =  - \frac{{\epsilon_{\rm m}^{\rm T}}}{{{\epsilon_{\rm d}}}} + \frac{{{k_{\rm sp}}(1 - \epsilon_{\rm m}^{\rm T})}}{{{k_{\rm lz}}}}\tanh \left(\tfrac{{{k_{\rm lz}}{a_{\rm m}}}}{2}\right),\\
\coth\left(\tfrac{{{k_{\rm sp}}{a_{\rm m}}}}{2}\right) =  - \frac{{\epsilon_{\rm m}^{\rm T}}}{{{\epsilon_{\rm d}}}} + \frac{{{k_{\rm sp}}(1- \epsilon_{\rm m}^{\rm T})}}{{{k_{\rm lz}}}}\coth\left(\tfrac{{{k_{\rm lz}}{a_{\rm m}}}}{2}\right),
\end{eqnarray}
\end{subequations}
where $k_{\rm sp}$ represents the SPP wavevector, and $k_{\rm lz}=({k_{\rm sp}^2}-{k_{L}^2})^{1/2}$. In the limit $a_{\rm m}\to0$, the dispersion equation of the even mode has no solution, but the odd mode always has one, even above $\omega_{\rm p}$.  Its dispersion is such that $k_{\rm{sp}}$ has $k_{\scriptscriptstyle \parallel}^c$ as an upper bound in the limit $a \to 0$. So we can now understand that it is this nonlocal `ceiling' for the single-layer SPP wavenumber that leads to a cutoff of $k_{\scriptscriptstyle \parallel}$ for the bulk modes of the metamaterial, as we saw in Fig.~\ref{fig2}.

\begin{figure}[h!]
\includegraphics[width=0.45\textwidth]{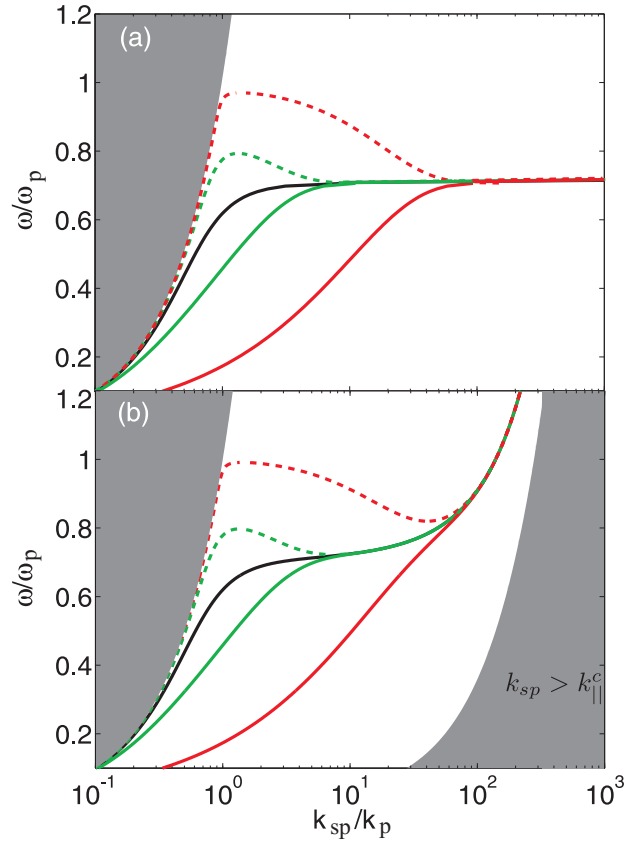}
\caption
{(Color online) Dispersion curves of the SPP mode supported by a single lossless Au layer with a thickness $a_{\rm m}$ in free space, in the (a) local-response aproximation, and (b) the hydrodynamic Drude model. Dashed and solid curves correspond to even and odd modes, respectively, with red curves for $a_{\rm m}=0.1\lambda_{\rm p}$, green for $a_{\rm m}=0.01\lambda_{\rm p}$, and black curves for $a_{\rm m}\to\infty$ (single-interface SPP). The gray areas are forbidden regions for the SPP modes, with light cones on the left.}
\label{fig3}
\end{figure}

In Fig.~\ref{fig3} we analyze numerically the effect of retardation on the SPP dispersion of a single Au layer in free space, for local and nonlocal response. With retardation, near the light cone also even-mode solutions exist. Only for nonlocal response do we find modes above $\omega_{\rm p}$. Again we find that nonlocal response gives rise to a forbidden region $k_{\rm sp}>k_{\scriptscriptstyle \parallel}^c$ for the odd SPP mode, see Fig.~\ref{fig3}(b). By contrast, in Fig.~\ref{fig3}(a) for the LRA, both even and odd modes have finite-frequency solutions with $k_{\rm sp}$ approaching infinity, which leads to the characteristic hyperbolic curve of the HMM that extends to infinitely large wavevectors.
%

\section{Local Density of States}
The discussed dramatic modification of the metamaterial dispersion  due to nonlocal response will also strongly affect the broadband super-singularity known to occur in the local-response LDOS, as we shall see. In general, the LDOS is proportional to the spontaneous-emission rate averaged over all solid angles, and defined as
\begin{equation}
{\rm LDOS}({\mathbf{r}_0},\omega) = -\frac{{2{k_0}}}{{3\pi c}}{\rm Tr}\bigl\{ {{\mathop{\rm Im}\nolimits} \left[ \bfsfG({\mathbf{r}_0},{\mathbf{r}_0},\omega)\right]} \bigl\},
\label{eqldos}
\end{equation}
where $\bfsfG$ is the dyadic Green function of the medium and ${\mathbf{r}_0}$ the position of the emitter. The Green function $\bfsfG$ is defined by
\begin{equation}
-\boldsymbol\nabla \times \boldsymbol\nabla  \times \bfsfG (\mathbf r,\mathbf r') +k_0^2  \int\mbox{d}{\bf r}_{1}\,\epsilon({\bf r},{\bf r}_{1}) \bfsfG({\bf r}_{1},{\bf r'})
= {\mbox{\sffamily\bfseries{I}}}\delta(\mathbf r-\mathbf r'),
\end{equation}
where ${\mbox{\sffamily\bfseries{I}}}$ represents the unit dyad, and $\epsilon$ represents the dielectric function, which is a position-dependent delta function for the local dielectric medium, and a tensorial nonlocal operator defined by Eq.~(\ref{hdm}) for the metal.
For the multilayered HMM, $\bfsfG$ can be decoupled into separate contributions from TM and TE modes. TM modes support the hyperbolic dispersion curve, and greatly dominate the LDOS, so we will neglect the TE contribution to the LDOS.

If we first neglect loss, then only radiative modes contribute to the LDOS. For an electric dipole with moment $\bm{\mu}$, the contribution to the LDOS of a single radiative mode is proportional to $|\bm{\mu}\cdot{\mathbf{a}_\mathbf{k}}(\mathbf{r}_0)|^2/|{\nabla_\mathbf{k}}\omega |$, where $\mathbf{a}_\mathbf{k}$ is the properly normalized mode function.\cite{Sanchez:1995}
In the LRA, for the limiting case of $a\to0$, the single-mode contribution to the LDOS scales linearly in $k$ as $k_{\scriptscriptstyle \parallel}$ and $k_z$ tend to infinity. This results in a diverging radiative LDOS, the broadband LDOS supersingularity of hyperbolic media.

Let us now consider the LDOS in the HDM instead. If we again take the limit $a\to0$, and let  $k_{\scriptscriptstyle \parallel}$ tend to $k_{\scriptscriptstyle \parallel}^c$ and $k_z$ to infinity, then this time the single-mode contribution to the LDOS scales as $1/{k_z}^2$, which we derived using the effective parameters of Eq.~(\ref{eq4}). Radiative modes with large wavenumbers are therefore  negligibly excited. As a main result of this paper, we consequently find that in the HDM the radiative LDOS converges to a finite value as $a\to0$, even though the integration area in k-space diverges. We find the numerically exact value and its analytical approximation
\begin{equation}
\mathrm{LDOS}(\omega)=\frac{{{\omega ^2}}}{{{6\pi^2\beta ^3}}}\eta,
\label{eq6}
\end{equation}
where
\begin{equation}
\eta =\frac{1}{\sqrt{\epsilon_{zz}^{\rm d}}}\int_{\theta_{0}}^{\pi /2}\mbox{d}\theta \frac{\cos^{2}\theta+(\epsilon_{\scriptscriptstyle \parallel}^{\rm d}/\epsilon_{zz}^{\rm d})[\sin^{2}\theta -\epsilon_{\scriptscriptstyle \parallel}^{\rm loc}/\epsilon_{\scriptscriptstyle \parallel}^{\rm d}]}{\sqrt{\sin^{2}\theta-\epsilon_{\scriptscriptstyle \parallel}^{\rm loc}/\epsilon_{\scriptscriptstyle \parallel}^{\rm d}}}
\label{eta}
\end{equation}
with $\theta_{0}$ equal to $\arcsin(\epsilon_{\scriptscriptstyle \parallel}^{\rm loc}/\epsilon_{\scriptscriptstyle \parallel}^{\rm d})$ for $\epsilon_{\scriptscriptstyle \parallel}^{\rm{loc}}>0$ and vanishing otherwise. The derivations leading to Eq.~(\ref{eq6}) are presented in Appendix~C. As illustrated below, Eq.~(\ref{eq6}) entails that nonlocal response leads to a large upper bound to the radiative LDOS of the HMM, proportional to $\omega^2/v_{\rm F}^3$. This exceeds the free-space radiative LDOS approximately by $c^3/v_{\rm F}^3$, which is of order $10^7$ for most metals.

When taking metallic Drude loss into account, then the LDOS has contributions both from radiative modes and from nonradiative quenching, the latter due to loss. For the limiting case of $a\to0$, we already discussed that  $\epsilon_{\scriptscriptstyle \parallel}^{\rm{nloc}}$ tends to $\epsilon_{\scriptscriptstyle \parallel}^{\rm d}$, see Eq.~(\ref{eq4}). For large wavevectors $k_{\scriptscriptstyle \parallel}$ also the other component $\epsilon_{zz}^{\rm{nloc}}$ tends to $\epsilon_{zz}^{\rm d}$. Thus, to the extent that $\epsilon_{\rm d}$ is lossless, the evanescent mode with large $k_{\scriptscriptstyle \parallel}$ does not contribute to the nonradiative LDOS, which therefore stays finite. As a result, the total LDOS containing both radiative and nonradiative contributions in the HDM converges as $a\to0$. In the low-loss case, where the radiation LDOS is dominant, Eq.~(\ref{eq6}) is an accurate expression of the total LDOS, as we verify by numerically exact simulation below.

We calculate the LDOS numerically exactly by merging two methods: the local-response transfer matrix method by Toma\v{s} to calculate the Green function of arbitrary multilayer media~\cite{Tomas:1995},
and the aforementioned HDM extension of the transfer matrix method.\cite{Mochan:1987} The details can be found in Appendix D.


Figure~\ref{fig4}(a) depicts the LDOS enhancement, defined as the ratio between LDOS in the HMM and in free space, as a function of the periodicity $a$. Clearly, in the HDM the LDOS converges to a finite value as $a\to0$. This proves that both the radiative and nonradiative LDOS in the HDM are finite.
By contrast, in the LRA the LDOS diverges as $1/a^3$, where the nonradiative LDOS has a dominant contribution.\cite{Jacob:2009,Barnett:1996,Chang:2006,Sun:2007}
The HDM ceases to be valid for $a < \lambda_{\rm F}$, but the LDOS enhancement value in the limit $a\to0$ is a useful upper bound.

We also calculated the LDOS for the lossless case [not shown Fig.~\ref{fig4}(a)], and we find smaller values for the LDOS owing to the missing nonradiative contribution, but the same trend for $a\to0$. This proves that nonlocal response rather than loss is responsible for removing the singularity of the radiative LDOS.
In Fig.~\ref{fig4}(b) we compare the limiting LDOS for the numerically exact method in the lossy case with the lossless analytical approximation of Eq.~(\ref{eq6}), when artificially varying the Fermi velocity. The value from the exact method is only larger than that from Eq.~(\ref{eq6}) by around $6\%$. We attribute the small difference in LDOS to the nonradiative LDOS due to the Drude loss. Thus, the loss acts as a small perturbation to the radiative LDOS.

\begin{figure}[h!]
\includegraphics[width=0.48\textwidth]{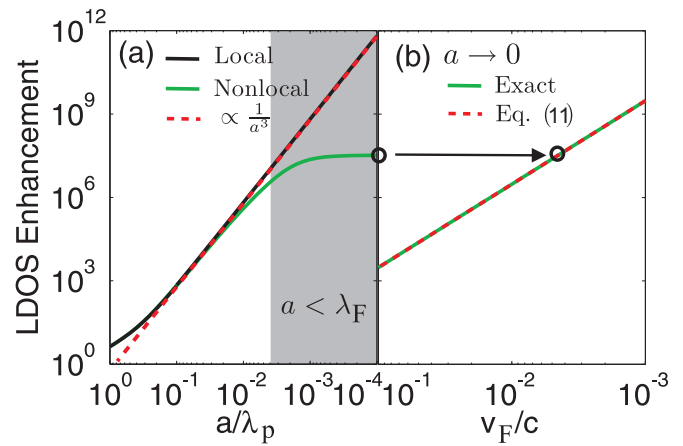}
\caption
{(Color online) (a) LDOS versus the periodicity $a$ at the center of the free space layer of the hyperbolic metamaterial for $\omega=0.2\omega_{\rm p}$. (b) The $a\to 0$ limiting value of the LDOS in the hydrodynamic Drude model as function of $v_{\rm F}/c$. Parameters of the hyperbolic metamaterial as in Fig.~\ref{fig2}.}
\label{fig4}
\end{figure}
%

\section{Discussion and conclusions}
For finite-sized unit cells, small loss gives rises to a regular perturbation of the radiative LDOS, both in the local and in the nonlocal response theories. However, the theories start to differ dramatically in the limit of infinitely small unit cells. In particular, in the nonlocal hydrodynamic Drude model the small variation of the radiative LDOS with small loss is quite different from the previously found radiative LDOS scaling with loss in the local theory as $\gamma^{-3/2}$ for infinitely small unit cells.\cite{Jacob:2009}

The small increase of the total LDOS due to loss is also quite different from spontaneous-emission rates of a point emitter inside a homogeneous absorbing medium, where the loss induces non-radiative quenching that can dramatically decrease the radiative decay efficiency.\cite{Jacob:2009,Barnett:1996,Chang:2006,Sun:2007}
In this sense, the nonlocal response regularizes the singularity not only of the radiative but also of the nonradiative LDOS of a lossy HMM.
One can interpret this finite nonradiative LDOS as due to a nonlocal screening of the electron scattering loss.\cite{Ford:1984} In a certain high wave-vector region the reverse can also occur, namely the enhancement of the nonradiative LDOS by the nonlocal response, when not only taking Drude loss into account, as we do here, but also electron-hole pair absorption. By neglecting any dielectric response of the metal apart from the (hydrodynamic) Drude response, we underestimate the nonradiative LDOS of real metals. However, the important conclusion that the nonlocal response removes the singularity of nonradiative LDOS is still valid.\cite{Ford:1984}

In conclusion, we have shown that the hydrodynamic Drude model gives closed non-hyperbolic dispersion relations for hyperbolic metamaterials, with a fundamental wavevector cutoff $\propto \omega/v_{\rm F}$. These effective dispersion relations have hyperbolic limits for small wavevectors, but the precise hyperbola  depends on the subwavelength size of the unit cell, contrary to consensus based on the local-response approximation. We find that the hydrodynamic model regularizes the broadband super-singularity of the radiative LDOS, and provides a large physical upper bound proportional to $\omega^2/v_{\rm F}^3$. In practice, considering the finite values of $a$ and $D$, i.e., the finite
sizes of the unit-cell and the emitter, we usually have $1/a<1/D<\omega/v_{\rm F}$. This indicates that the size effects have a dominant role in limiting the LDOS enhancement. Thus, under an upper bound set up by the nonlocal response, hyperbolic metamaterials have a plenty of room for improvement in boosting light-matter interactions by decreasing the sizes of the unit-cell and the emitter.

\section*{Aknowledgments}
We thank S. Raza for stimulating discussions. This work was financially supported by an H.~C. {\O}rsted Fellowship (W.Y.).

\appendix

\section{Boundary Conditions}
Since the hydrodynamic dynamics allows the excitation of longitudinal waves, the unambiguous solution of the nonlocal-response dynamics requires additional boundary conditions (ABCs), complementing the Maxwell boundary conditions. As is well known, the Maxwell boundary conditions are a consequence of Maxwell's equations themselves, in the sense that the derivation of the boundary conditions only involves Maxwell's equations plus mathematics (the Gauss and Stokes theorems).
Quite analogously, ABC's are not a matter of choice but can be derived from the (linearized) hydrodynamic equations, at least for a given equilibrium free-electron density profile $n_{0}$.\cite{Jewsbury:1981a,Raza:2011} When assuming a simple zero-to-nonzero step  profile of $n_{0}$ at the dielectric-metal interfaces, this unambiguously leads to one and only one required ABC, namely the continuity of the normal component of the free-electron current ${\bf J}$.\cite{Jewsbury:1981a,Raza:2011}

Let us now write the relative permittivity of the dielectric medium as  $\epsilon_{\rm d}$, and the dielectric response of the metal as $\epsilon_{\rm m}(\omega)$. We assume that $\epsilon_{\rm m}(\omega)$  is given by the sum of a nonlocal hydrodynamic Drude free-electron response plus $\epsilon_{\rm m}^{\rm other}(\omega)$, the latter describing the remaining dielectric response of the metal.
Since one of the Maxwell boundary conditions is the conservation of the normal component of the displacement field, the ABC is equivalent to the condition
\begin{equation}
\epsilon_{\rm m}^{\rm other}\mathbf E_{\rm m}\cdot\hat n=\epsilon_{\rm d} \mathbf E_{\rm d}\cdot\hat n,
\label{ABC1}
\end{equation}
where $\mathbf E_{\rm m,d}$ represent the electric fields in the metal and dielectric, respectively, and $\hat n$ is the unit vector normal to the boundary.  From Eq.~(\ref{ABC1}), we see that the normal electric field is discontinuous across the boundary when $\epsilon_{m}^{\rm other}\ne \epsilon_d$. A jump in the electric field occurs due to the surface charge produced by polarization of the {\em bound} electrons both in the dielectric and in the metal.

We discuss the ABC in some detail, because  Moch{\' a}n {\em et al.}~\cite{Mochan:1987}, whose pioneering transfer matrix method we employ here,  and also recently Cirac{\` i} {\em et al.}\cite{Ciraci:2012b} used instead the continuity of the normal component of the electric field as the ABC,
\begin{equation}
\mathbf E_{\rm m}\cdot\hat n=\mathbf E_{\rm d}\cdot\hat n,
\label{ABC2}
\end{equation}
or equivalently the continuity of the normal component of the displacement current. There is no derivation of the latter ABC in Refs.~\cite{Mochan:1987,Ciraci:2012b}. It happens to be only correct, in agreement with Eq.~(\ref{ABC1}), if $\epsilon_{\rm m}^{\rm other}=\varepsilon_{\rm d}$, for example in case the dielectric is vacuum ($\varepsilon_{\rm d}=1$) and the metal is a pure Drude metal ($\epsilon_{\rm m}^{\rm other}=1$).  Moch{\' a}n {\em et al.}~\cite{Mochan:1987} applied their ABC for simplicity and write that they thereby ignore the discontinuity of the electric field, due to the accumulation at the surface of bound charges. Our main point is here that without additional complication the correct ABC can be implemented, and that many physical predictions of the hydrodynamic Drude model are sensitive to implementing the ABC correctly.

To understand the ABC physically, recall that in the HDM the dynamics of the free electrons is described by the equation of motion
\begin{equation}
{m_e}\left[ {\frac{{\partial \mathbf v}}{{\partial t}} + \mathbf v \cdot\boldsymbol \nabla\mathbf v} \right] = - \frac{{\boldsymbol\nabla {p_{\deg }}}}{n} + e\left( {\mathbf E + \mathbf v \times \mathbf B} \right),
\label{emotion}
\end{equation}
where $p_{\deg}$ is the pressure from the ground state energy of the degenerate quantum Fermi gas, and $n$ is the free-electron density. The pressure force $-{{\boldsymbol\nabla {p_{\deg }}}}/{n}$ $\propto  - \boldsymbol\nabla n/n$ drives the free electrons diffusing from the high density region to the low density region. It is this force that prevents the free-electron charge from accumulating on the boundary surface, since the existence of a free-electron surface charge would cause an infinitely large pressure force, which is unphysical. The nonexistence of the surface free-electron charge indicates that the free-electron current should be continuous across the boundary, as the ABC~(\ref{ABC1}) indeed describes.
By contrast, in the ABC of Eq.~(\ref{ABC2}), there exists no surface charges at all. This indicates that the pressure force somehow smears out not only the surface free-electron charge in the metal but also the surface polarization charges  in both the metal and the dielectric. However, one cannot expect the smearing out of the surface polarization charges in the HDM, since the pressure force only acts on the free electrons in the metal. In this sense, the ABC of Eq.~(\ref{ABC2}) is not consistent with the assumed dynamics and thus not physically sound.

There is another perhaps simpler argument, a consistency check that confirms that the  ABC of Eq.~(\ref{ABC2}) is more  problematic. 
Assume there is a thin free-space layer with subwavelength thickness $\delta$ between the nonlocal metal and the local dielectric medium. At the boundary between the metal and free space, the ABC of  Eq.~(\ref{ABC2}) gives $\mathbf E_{\rm m}\cdot\hat n=\mathbf E_{\rm f}\cdot\hat n$, where $\mathbf E_{\rm f}$ represents the electric field in the free-space layer. At the boundary between free space and the dielectric medium, we have $\mathbf E_{\rm f}\cdot\hat n=\epsilon_{\rm d}\mathbf E_{\rm d}\cdot\hat n$ by the standard Maxwell boundary condition of the continuity of the normal component of the displacement field. In the limit of an infinitely thin free-space middle layer ($\delta\to 0$), the three-layer system essentially becomes the two-layer system where the metal and the dielectric medium touch, and for which we find $\mathbf E_{\rm m}\cdot\hat n=\epsilon_{\rm d}\mathbf E_{\rm d}\cdot\hat n$ by combining the previous two identities. However, this contradicts with Eq.~(\ref{ABC2}) for the metal-dielectric interface. Thus, the ABC of Eq.~(\ref{ABC2}) can not be applied consistently. For the ABC of Eq.~(\ref{ABC1}), we obtain instead consistent results when following the above thin-layer argument.

\section{Effective material parameters of hyperbolic metamaterial}
When the unit cell has a thickness $a$ that is much smaller than an optical wavelength $\lambda_{0}$, then the optical properties of such an infinite multilayer structure can be macroscopically described by a diagonal effective dielectric tensor $\boldsymbol \epsilon=\rm{diag}[\epsilon_{\scriptscriptstyle \parallel},\epsilon_{\scriptscriptstyle \parallel},\epsilon_{zz}]$ with tensor components
\begin{equation}\label{averaging}
\epsilon_{\scriptscriptstyle \parallel}=\frac{\langle D_{x,y}\rangle}{\langle E_{x,y}\rangle},\;\epsilon_{zz}=\frac{\langle D_{z}\rangle}{\langle E_{z}\rangle},
\end{equation}
and where $\langle \ldots\rangle$ denotes spatial averaging over a unit-cell.

The unit cell can be chosen symmetric, identical for left- and right-traveling waves. Consider a unit-cell positioned at $-a/2<z<a/2$, with the metal layer at $-a_{\rm m}/2<z<a_{\rm m}/2$, which is symmetric in $z=0$. The total fields in such a unit cell are generated by waves incident both from the left (``${\rm l}$'') and from the right ``${\rm r}$'', and the average fields can be split into two terms, $\langle\mathbf{E}\rangle=\langle\mathbf{E}\rangle_{\rm l}+\langle\mathbf{E}\rangle_{\rm r}$. However, by symmetry of the unit cell it follows that $\boldsymbol\epsilon=\langle\mathbf{D}\rangle/\langle\mathbf{E}\rangle=
\langle\mathbf{D}\rangle_{\rm l}/\langle\mathbf{E}\rangle_{\rm l}=\langle\mathbf{D}\rangle_{\rm r}/\langle\mathbf{E}\rangle_{\rm r}$. To obtain the effective material parameters, we can simply replace the average fields in the periodic structure by the average fields in a single unit cell.

Before spatially averaging the fields, we first need to find them as solutions of Maxwell's equations. We focus solely on TM-polarized waves since the hyperbolic dispersion occurs for those waves only. Since $k_0a_{\rm m}\ll1$, we can make the quasi-static approximation, where $\mathbf{E}=-\nabla\phi$. Consider an incident electric field with the electric potential $\phi={\rm{exp}}(ik_{\scriptscriptstyle \parallel}x-k_{\scriptscriptstyle \parallel}z)$. The electric potential in the whole system can then be written as
\begin{eqnarray}
&\;&{\phi _1} = \exp (i{k_{\scriptscriptstyle \parallel}}x)\left[ {\exp ( - {k_{\scriptscriptstyle \parallel}}z) + r\exp ({k_{\scriptscriptstyle \parallel}}z)} \right],\nonumber\\
&\;&\phi _2^{\rm T} = \exp (i{k_{\scriptscriptstyle \parallel}}x)\left[ {{A_1}\exp ( - {k_{\scriptscriptstyle \parallel}}z) + {A_2}\exp ({k_{\scriptscriptstyle \parallel}}z)} \right],\nonumber\\
&\;&\phi _2^{\rm L} = \exp (i{k_{\scriptscriptstyle \parallel}}x)\left[ {{B_1}\exp ( - {k_{Lz}}z) + {B_2}\exp ({k_{Lz}}z)} \right],\nonumber\\
&\;&{\phi _3} = t\exp (i{k_{\scriptscriptstyle \parallel}}x)\exp ( - {k_{\scriptscriptstyle \parallel}}z),\nonumber\\
\end{eqnarray}
where $k_{Lz}=\sqrt{k_{\scriptscriptstyle \parallel}^2-k_{\rm L}^2}$. By matching boundary conditions at the two metal-dielectric interfaces, the above equations can be solved. After obtaining the field distributions, we can average the fields from $-a/2<z<a/2$ to obtain the effective material parameters using Eq.~(\ref{averaging}).

First consider the case in which the metal layers are much thicker than the wavelength of the longitudinal waves ($k_{{\rm L}z}a_{\rm m}\gg1$), but where the unit cell is much thinner than an optical wavelength. Here we expect an effective (homogenized) description to apply and nonlocal response to be negligible. Following the above described scheme, the effective material parameters, to the zeroth-order in the small parameters $k_{\scriptscriptstyle \parallel}a$ and $1/(k_{{\rm L}z}a_{\rm m})$, are found to be
\begin{equation}
\epsilon_{zz}^{\rm loc}=\frac{1}{{\frac{{{f_{\rm d}}}}{{{\epsilon _{\rm d}}}} + \frac{{{f_{\rm m}}}}{{{\epsilon _{\rm m}^{\rm T}}}}}},\qquad\;\epsilon_{\scriptscriptstyle \parallel}^{\rm loc}=f_{\rm d}\epsilon_{\rm d}+f_{\rm m}\epsilon_{\rm m}^{\rm T},
\label{eq1}
\end{equation}
in terms of the filling factors $f_{\rm d}=a_{\rm d}/a$ and $f_{\rm m}=a_{\rm m}/a$. Indeed, the effective material parameters are just as what one would find in the local response approximation (LRA).

Second, we consider the limiting case of $a_{\rm m}\to0$ with $k_{{\rm L}z}a_{\rm m}\ll 1$, where the nonlocal response is extremely strong. As before, we keep the filling fractions $f_{\rm m,d}$ constant when taking the limit.  The effective material parameters, to zeroth order in both  $k_{\scriptscriptstyle \parallel}a$ and $k_{{\rm L}z}a_{\rm m}$, become
\begin{equation}
{\epsilon_{zz}^{\rm{nloc}}}=\epsilon_{zz}^{\rm{d}},\;\;\;\epsilon_{\scriptscriptstyle \parallel}^{\rm{nloc}} = \epsilon_{\scriptscriptstyle \parallel}^{\rm{d}}\frac{{k_L^2\epsilon_{\scriptscriptstyle \parallel}^{\rm{loc}}/\epsilon_{\scriptscriptstyle \parallel}^{\rm{d}} - k_{\scriptscriptstyle \parallel}^2{\epsilon_{\rm m}^{\rm T}}}}{{k_L^2 - k_{\scriptscriptstyle \parallel}^2{\epsilon_{\rm m}^{\rm T}}}},
\label{eq2}
\end{equation}
Eq.~(\ref{eq2}) characterizes how the nonlocal response can modify the effective material parameters to the largest extent.

\section{Limiting LDOS of hyperbolic metamaterials}
Here we provide the calculation details of the LDOS in the hydrodynamic Drude model in the limit of infinitely small unit cells ($a\to0$). We employ the effective material parameters derived in Eq.~(\ref{eq2}). Since the by far dominant contribution to the LDOS stems from TM waves, we will neglect the TE contribution. In $k$-space, the diagonal components of $\bfsfG$, in an effective medium with material parameters expressed in Eq.~(\ref{eq2}) for TM polarization, are found to be
\begin{eqnarray}
\sfG_{{\rm TM},jj}^\mathbf{k} &=& \frac{1}{{k_{\scriptscriptstyle \parallel}^2\epsilon _{\scriptscriptstyle \parallel}^{\rm{nloc}}}}\frac{{k_x^2[1-k_{\scriptscriptstyle \parallel}^2/(k_0^2\epsilon _{zz}^{\rm{nloc}})]}}{{k_0^2 - k_{\scriptscriptstyle \parallel}^2/\epsilon _{zz}^{\rm{nloc}} - k_z^2/\epsilon _{\scriptscriptstyle \parallel}^{\rm{nloc}}}}\quad\mbox{for}\; j=x,y,\nonumber\\
\sfG_{{\rm TM},zz}^\mathbf{k} & = &\frac{1}{{\epsilon _{zz}^{\rm{nloc}}}}\frac{{1-k_z^2/(k_0^2\epsilon _{\scriptscriptstyle \parallel}^{\rm{nloc}})}}{{k_0^2 - k_{\scriptscriptstyle \parallel}^2/\epsilon _{zz}^{\rm{nloc}} - k_z^2/\epsilon _{\scriptscriptstyle \parallel}^{\rm{nloc}}}}.
\end{eqnarray}
When inserting these diagonal components for the Green tensor into expression Eq.~(\ref{eqldos}) for the LDOS, we obtain
\begin{widetext}
\begin{eqnarray}
\mathop {\lim }\limits_{a \to 0} {\rm LDOS}&=&-\frac{{2{k_0}}}{{3\pi c}}\frac{1}{(2\pi)^3}{\mathop{\rm Im}\nolimits} \int\mbox{d}^{3}{\mathbf k}\sum_{j = x,y,z}
 \sfG_{{\rm TM}, jj}^\mathbf{k}\nonumber\\
&=&-\frac{{ {k_0}}}{{6\pi^2c}}{\mathop{\rm Re}\nolimits} \left[ {\int_0^\infty  {\mbox{d}{k_{\scriptscriptstyle \parallel}}\frac{k_{\scriptscriptstyle \parallel}-\frac{k_{\scriptscriptstyle \parallel}^3}{k_0^2}\frac{1-\epsilon _{\scriptscriptstyle \parallel}^{\rm nloc}/\epsilon _{zz}^{\rm nloc}}{\epsilon_{zz}^{\rm nloc}}}{{\sqrt {k_0^2\epsilon _{\scriptscriptstyle \parallel}^{\rm{nloc}} - k_{\scriptscriptstyle \parallel}^2\epsilon _{\scriptscriptstyle \parallel}^{\rm{nloc}}/\epsilon _{zz}^{\rm{nloc}}} }}} } \right]\nonumber\\
&\approx&\frac{{ {k_0}}}{{6\pi^2c}}{\mathop{\rm Re}\nolimits} \left[ {\int_0^\infty  {\mbox{d}{k_{\scriptscriptstyle \parallel}}\frac{k_{\scriptscriptstyle \parallel}^2}{k_0^2}\frac{\frac{1-\epsilon_{\scriptscriptstyle \parallel}^{\rm nloc}/\epsilon_{zz}^{\rm nloc}}{\epsilon_{zz}^{\rm nloc}}}
{{\sqrt {-\epsilon _{\scriptscriptstyle \parallel}^{\rm{nloc}}/\epsilon _{zz}^{\rm{nloc}}} }}} } \right]\nonumber\\
&=&\frac{{{\omega ^2}}}{{{6\pi^2\beta ^3}}}\eta,
\label{eqldosl}
\end{eqnarray}
\end{widetext}
where $\eta$ is expressed in Eq. (\ref{eta}). In the derivation, we neglected losses in the metal. To arrive at the second line of Eq. (\ref{eqldosl}), we use the principal-value identity $\mathop {\lim }\limits_{\Delta  \downarrow {0}} \frac{1}{{x \pm i\Delta }} = P\frac{1}{x} \pm i\pi \delta (x)$. The final identity then follows immediately by inserting the $a \to 0$ limiting expressions for $\varepsilon_{zz}^{\rm nloc}$ and $\varepsilon_{\scriptscriptstyle \parallel}^{\rm nloc}$ given in Eq.~(\ref{eq2}).

\section{Green Function of hyperbolic metamaterial}
Consider an emitter positioned in the dielectric layer of the HMM. The HMM can be divided into three regions: (1) the central dielectric layer where the emitter is located; (2) the left semi-infinite HMM; (3) the right semi-infinite HMM. The distance between the emitter and the left (right) boundary of the dielectric layer is $z_l$ ($z_r$).
The Green function $\bfsfG$ in the central layer could be separated into two terms
\begin{eqnarray}
\bfsfG(\mathbf{r},\mathbf{r}_0)=\bfsfG_{\rm d}(\mathbf{r},\mathbf{r}_0)+\bfsfG_{\rm s}(\mathbf{r},\mathbf{r}_0),
\end{eqnarray}
where $\bfsfG_{\rm d}$ represents the Green's function for the emitter in the homogenous dielectric medium, while $\bfsfG_{\rm s}$ represents the Green's functions owing to the scattering between central layer and the left and right semi-infinite HMM. In the plane wave basis, $\bfsfG_{\rm d}$ is expressed as \cite{Tomas:1995}
\begin{eqnarray}
{\bfsfG_{\rm d}}(\mathbf{r},{\mathbf{r_0}})&=& -\frac{\delta (z - {z_0})}{{{k_{\rm d}}^2}}\hat z\hat z\int {d^2\mathbf{k_{\scriptscriptstyle \parallel}}\exp \left[ {i\mathbf{{k_{\scriptscriptstyle \parallel}}} \cdot (\mathbf{r_{\scriptscriptstyle \parallel}} - \mathbf r_{0\scriptscriptstyle \parallel})} \right]} \nonumber\\
&+&\frac{i}{{8{\pi ^2}}}\int d^2{\mathbf{k_{\scriptscriptstyle \parallel}}}{\frac{\left[ {{\mathbf e_{\rm TE}}{\mathbf e_{\rm TE}} + {\mathbf e_{\rm TM}^{\pm}}{\mathbf e_{\rm TM}}^{\pm}} \right]}{{{k_z}}}}\nonumber\\
&\;& {\exp \left[ {i\mathbf k_{\scriptscriptstyle \parallel} \cdot (\mathbf r_{\scriptscriptstyle \parallel} - \mathbf r_{0\scriptscriptstyle \parallel})}+ik_z|z-z_0|\right]},
\label{eq3}
\end{eqnarray}
with
\begin{eqnarray}
\mathbf e_{\rm TE}&=&\frac{\mathbf k_{\scriptscriptstyle \parallel}}{k_{\scriptscriptstyle \parallel}}\times \hat z,\nonumber\\
\mathbf e_{\rm TM}^{\pm}&=&\frac{\mathbf k_{\scriptscriptstyle \parallel}\pm k_z\hat z}{k_d}\times\mathbf e_{\rm TE},
\end{eqnarray}
where $\mathbf{k}_{\scriptscriptstyle \parallel}=k_x\hat x+k_y\hat y$, $k_{\rm d}=\omega\sqrt{\epsilon_{\rm d}}/c$,
$k_z^2+k_{\scriptscriptstyle \parallel}^2=k_{\rm d}^2$, and $ \mathbf e_{\rm TM}^{\pm}$ correspond to $z>z_0$ and
$z<z_0$, respectively. The terms containing $ \mathbf e_{\rm TE}$ and $\mathbf e_{\rm TM}$ represent TE and TM waves, respectively.

The scattering part $\bfsfG_{\rm s}$ of the Green function is expressed as
\begin{widetext}
\begin{eqnarray}
{\bfsfG}_{\rm s}(\mathbf{r},\mathbf{r_0}) &=& \frac{i}{{8{\pi ^2}}}\int {d\mathbf{k}_{\scriptscriptstyle \parallel}\frac{1}{{{k_z}}}\exp \left[ {i\mathbf{k}_{\scriptscriptstyle \parallel} \cdot (\mathbf{r}_{\scriptscriptstyle \parallel} - \mathbf{r}_{0\scriptscriptstyle \parallel})} \right]}\nonumber\\
&\;&\times \left\{ {\left[ {r_{{\rm TE}}^{ +  + }\mathbf e_{\rm TE}\mathbf e_{\rm TE} + r_{{\rm TM}}^{ +  + }\mathbf e_{\rm TM}^+\mathbf e_{\rm TM}^+} \right.} \right.
\left. {+{r_{{\rm TE}}^{ +  -}\mathbf e_{\rm TE}\mathbf e_{\rm TE} + r_{{\rm TM}}^{ +  - }\mathbf e_{\rm TM}^+\mathbf e_{\rm TM}^{-}}}\right]\exp \left[ {i{k_z}(z - {z_0})} \right]
\nonumber\\&\;&
+{\left[ {r_{{\rm TE}}^{ - + }\mathbf e_{\rm TE}\mathbf e_{\rm TE}+ r_{{\rm TM}}^{ -  + }}\mathbf e_{\rm TM}^{-}\mathbf e_{\rm TM}^{+} \right.}
\left. {+{r_{{\rm TE}}^{ -  -}\mathbf e_{\rm TE}\mathbf e_{\rm TE}+ r_{{\rm TM}}^{ -  - }\mathbf e_{\rm TM}^{-}\mathbf e_{\rm TM}^{-}}}\right]
\left.{ \exp \left[ {-i{k_z}(z - {z_0})} \right]}\right\},
\end{eqnarray}
\end{widetext}
where $r_{_{{\rm TE},{\rm TM}}}^{\pm\pm}$ is the reflection coefficient, in which the left superscript ``$\pm$'' represents the scattering wave in the $\pm\hat{z}$ direction, and the right superscript ``$\pm$'' represents the incident wave in the $\pm\hat {z}$ direction. The $r_{_{{\rm TE},{\rm TM}}}^{\pm\pm}$ are found to be
\begin{eqnarray}
&\;&r_{{\rm TE},{\rm TM}}^{ +  + } = r_{{\rm TE},{\rm TM}}^{ -  - }=\frac{{R_{{\rm TE},{\rm TM}}^2\exp (2i{k_z}a_{\rm d})}}{{1 - R_{\rm TE,\rm TM}^2\exp (2i{k_z}a_{\rm d})}},\nonumber\\
&\;&r_{{\rm TE},{\rm TM}}^{ -  + } = \frac{{{R_{\rm TE,\rm TM}}\exp (2i{k_z}{z_r})}}{{1 - R_{{\rm TE},{\rm TM}}^2\exp (2i{k_z}a_{\rm d})}},\nonumber\\
&\;&r_{{\rm TE},{\rm TM}}^{ +  - } = \frac{{{R_{\rm TE,\rm TM}}\exp (2i{k_z}{z_l})}}{{1 - R_{{\rm TE},{\rm TM}}^2\exp (2i{k_z}a_{\rm d})}},
\end{eqnarray}
where $R_{{\rm TE},{\rm TM}}$ denote the reflection between the dielectric medium and the semi-infinite HMM for TE and TM waves, respectively, $z_{r}$ ($z_{l}$) represents the distance between the point emitter and the right (left) semi-infinite HMM. $R_{{\rm TE},{\rm TM}}$ can be calculated by the transfer matrix method as demonstrated in
Ref.~\onlinecite{Mochan:1987}.


\end{document}